\input{aipcheck}
\documentclass[
    ,final            
  ]
  {aipproc}

\layoutstyle{8x11double}
\newcommand{\lsim}{\raisebox{-0.7ex}{$\stackrel{\textstyle <}{\sim}$ }}
\newcommand{\gsim}{\raisebox{-0.7ex}{$\stackrel{\textstyle >}{\sim}$ }}

\begin{document}

\title{Nuclear Physics from QCD : The Anticipated Impact of Exa-Scale 
Computing}
\author{Martin J. Savage}{
  address={Department of Physics, University of Washington, Seattle, 
WA 98195-1560.}
}

\begin{abstract}
I will discuss highlights in the progress that is being made toward 
calculating processes of importance in nuclear physics from QCD using
high performance computing. 
As exa-scale computing resources are expected to become available around 2017, 
I present current estimates of the computational resources required to
accomplish central goals of nuclear physics.
\end{abstract}

\maketitle

Nuclear physics is a vast and  rich field  whose phenomenology has been
explored for decades through intense experimental and theoretical
effort. However, there is still little understanding of the connection
to the underlying theory of the strong interactions,
Quantum Chromodynamics (QCD), and the basic building blocks of nature, 
quarks and gluons.
A key element in   
connecting QCD to nuclei is
determining how the properties of the
hadrons,
and their responses to external probes, can be described and predicted 
in terms of  quark and gluon distributions. 
This is the focus of both theoretical and experimental
efforts world-wide, and in particular, at the Jefferson Laboratory.  
At the next length scale up from single hadrons are
nuclei, which are systems of neutrons and
protons (and also hyperons) that are  
bound by the nuclear forces, 
and it is this area that traditionally defined the field of nuclear physics.
Nuclear physics also encompasses the study of matter under extreme conditions
of temperature and density.  This continues to be the subject of extensive
experimental efforts using heavy-ion collisions, 
such as produced at RHIC at Brookhaven National Laboratory and now
at the LHC at CERN.  
The range of densities that are explored in such collisions is not great, and
does not match those found in some stellar environments
such as in the late-stages of evolution of Type II Supernova - the
birthplace of the heavy-elements.
Nuclear physics also impacts our understanding of 
fundamental cosmological aspects of our universe.
Type I Supernova are used as
standard candles to `calibrate' the universe.
In fact, they 
provided the first measurement of the acceleration of the 
universe and presently
provide a good
constraint on the dark-energy-dark-matter composition of the universe.

During 2009, a series of workshops were held in the United States that 
established the scientific needs for exa-scale computational 
resources for the field of Nuclear Physics.
Just to set the scale - 1~Exaflop = 1000 Petaflops = $10^6$ Teraflops = $10^9$
Gigaflops.
Partner workshops were held in the areas of Biology, Basic Energy Sciences,
Climate Science, Fusion Energy, High Energy Physics, National Security, Nuclear
Energy, Architectures and Technology, and Cross-cutting areas, the details of
which can be found in Ref.~\cite{EXTREMESCALEWORKSHOPS:2009}.  
In 2010 the
output, conclusions, and recommendations from the panels at the  
workshops were synthesized 
into an overall report.
The nuclear physics panel determined that
exa-scale computing is required for the calculation,
 with quantifiable uncertainties,
of  systems and processes
involving strongly interacting matter over a large range of temperatures and
densities.

Due to the increasing 
computational resources becoming available to the field,  
progress  is being made toward a direct
connection between QCD and nuclear physics using the numerical technique of
Lattice QCD (LQCD).  
It has been known for the past four decades that QCD, 
together with the electroweak interactions,
underlies all of nuclear physics.
However, 
soon after the discovery of QCD it became apparent that
the complexities of the theory at strong coupling would hinder
analytic progress in understanding the properties of the simplest
hadrons, let alone the simplest features of the nuclear forces.
Wilson pointed the way to eventual
direct quantitative confirmation of the origins of nuclear physics by
formulating Lattice QCD (LQCD)~\cite{Wilson:1974sk}, a regularization 
and non-perturbative definition
of QCD
that is suitable for the intensive computational demands of solving
QCD in the infrared.
LQCD is a technique in which Euclidean space correlation
functions are calculated by a Monte-Carlo evaluation of the Euclidean
space path integral~\cite{Wilson:1974sk}.  The calculations are
performed in Euclidean space so that  field configurations
that have a large action are exponentially
suppressed.  This is in contrast with Minkowski space in which large
action contributions result in a complex phase which will average to
an exponentially small contribution with nearby configurations.  
Space-time is
discretized with the quarks residing on the lattice sites, and the
gluon fields residing on the links between lattice sites. 
The lattice spacing, the distance between adjacent lattice sites,
is required to be much smaller than the characteristic hadronic length
scale of the system under study. The effects of a finite lattice
spacing can be systematically removed by combining calculations of 
correlation functions at several lattice spacings with the low-energy
effective field theory (EFT) which explicitly includes  the 
discretization effects. 
This type of EFT is somewhat more complicated than its
continuum counterpart as it must reproduce matrix elements of the
Symanzik action constructed with higher dimension operators induced by
the discretization~\cite{Symanzik:1983gh}. While the action lacks
Lorentz invariance and rotational symmetry, it is constrained by
hypercubic symmetry.  As computers have finite memory and performance,
the lattice volumes are finite in all four space-time directions.
Generally, periodic boundary conditions (BC's) are imposed on the
fields in the space-directions (a three-dimensional torus), while
(anti-) periodic BC's are imposed on the (quark) gauge fields 
in the time-direction,
which in many cases is much larger than the space-directions 
(in order to approach the zero-temperature limit).
During the last five years, LQCD has emerged from a long
period of research and development to the
point where precise predictions for hadronic quantities are starting to 
be made. In particular, presently, fully-dynamical calculations with
near-exact chiral symmetry at finite lattice-spacing have become standard,
with lattice volumes of spatial extent $L \gsim 2.5~{\rm fm}$ and
with lattice spacings in the range $b\lsim 0.12~{\rm fm}$. 
It is still the norm that the light-quark
masses, $m_q$, are larger than those of nature, with typical pion
masses $m_\pi\lsim 300~{\rm MeV}$.   However, exploratory calculations at, and
near, the physical pion mass, $m_\pi\sim 139~{\rm MeV}$, have been performed, but
the lattice volumes at these quark masses
remain too small for the results to be meaningfully compared with nature.

\begin{figure}[!ht]
  \includegraphics[height=0.26\textheight]{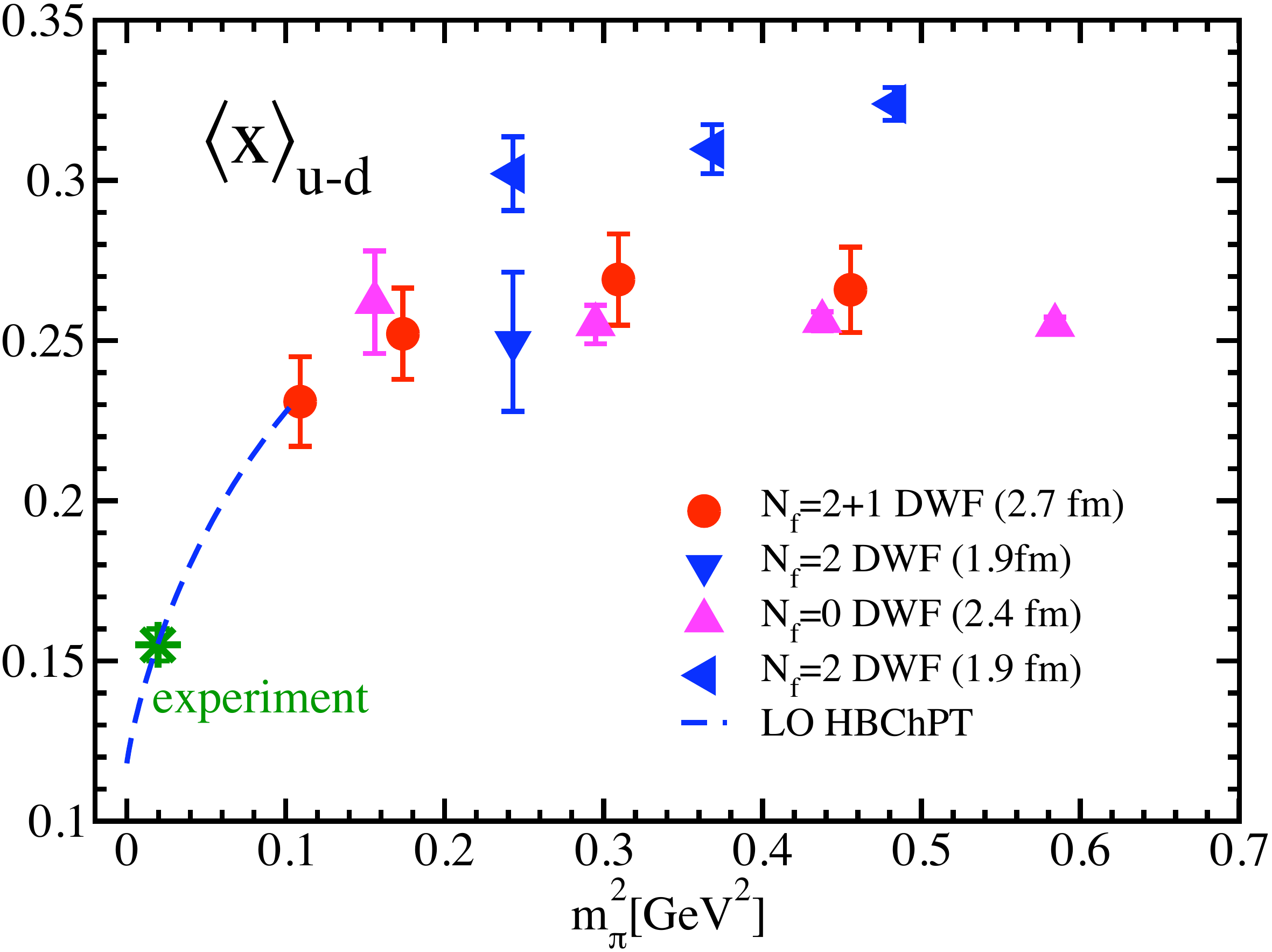}
  \caption{
LQCD calculations of the first moment of the isovector parton distribution in
the proton performed with the domain-wall fermion
discretization of the quark-fields~\protect\cite{Aoki:2010xg}.
I thank the authors for allowing me to reproduce this figure.
}
\label{fig:DWpdf}
\end{figure}

\begin{figure}[!ht]
  \includegraphics[height=0.24\textheight]{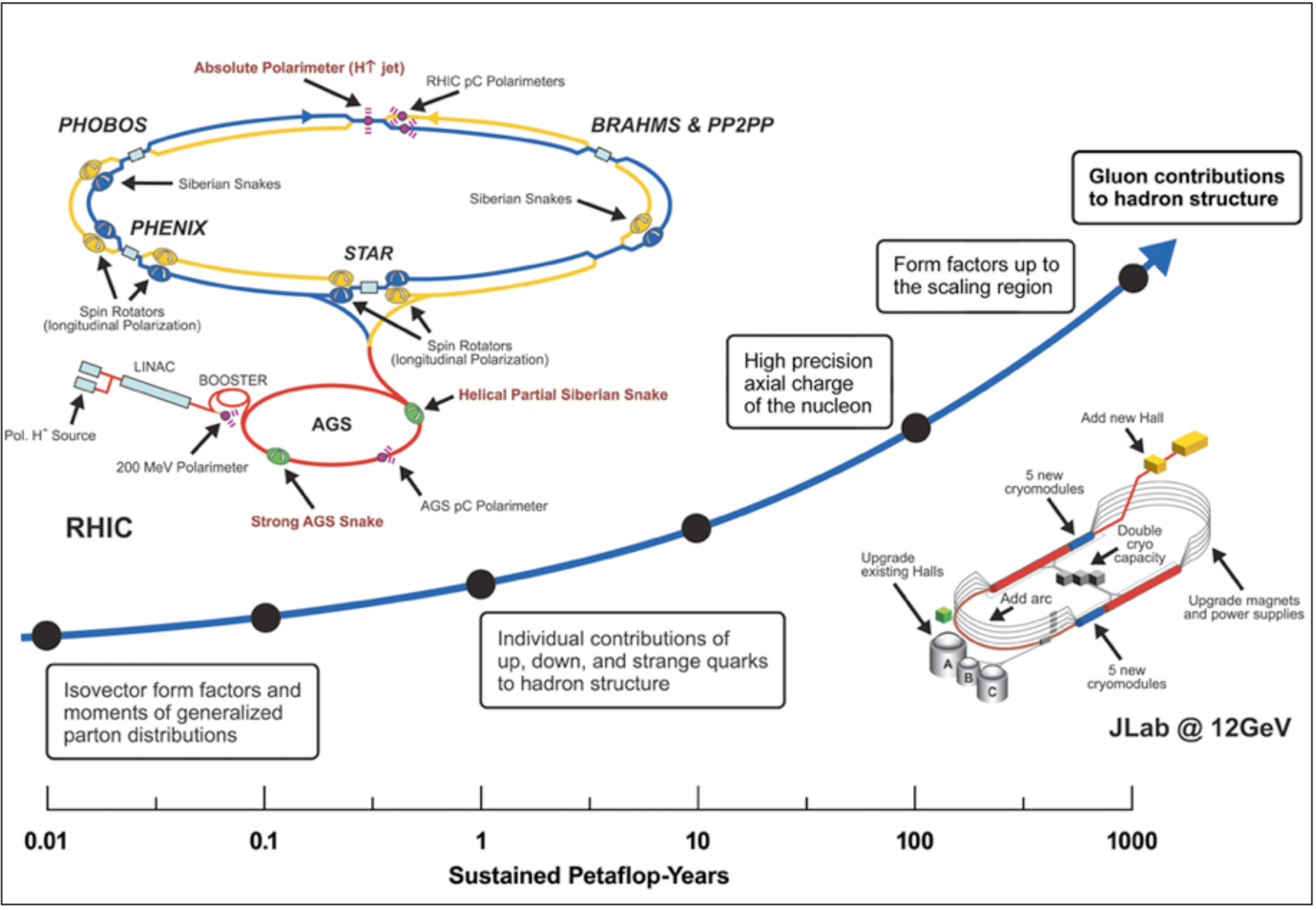}
  \caption{Estimates of the  computational resources required to calculate the
   structure of nucleons~\protect\cite{ExascaleReport:2009}.
}
\label{fig:makeaproton}
\end{figure}

Zero-temperature LQCD is a very active field.  This area is focused on the
determination of the properties of the hadrons, establishing the ground-state
masses and excitation spectrum, and  mapping out the parton 
and 
generalized parton distributions (as an example, see figure~\ref{fig:DWpdf}).  
This program is closely tied to the parallel
experimental efforts, in particular at the Jefferson laboratory, and has the 
exploration of exotic hadrons as a priority.  Exotic hadrons 
are ones that are not predicted by the 
most naive quark-model, but follow naturally from QCD - states with
interpolating operators that can be constructed only by including the gluon
field.  
A main focus of the efforts in fundamental symmetries is the prediction of the 
neutron electric-dipole moment resulting from a  
non-zero value of $\theta$, and from higher dimension
operators that may also contribute.
An estimate of the computational resources
required to calculate the structure of the nucleon can be seen in figure~\ref{fig:makeaproton}.
\vskip 0.1in
\vskip 0.1in

\begin{figure}[!ht]
  \includegraphics[height=0.22\textheight]{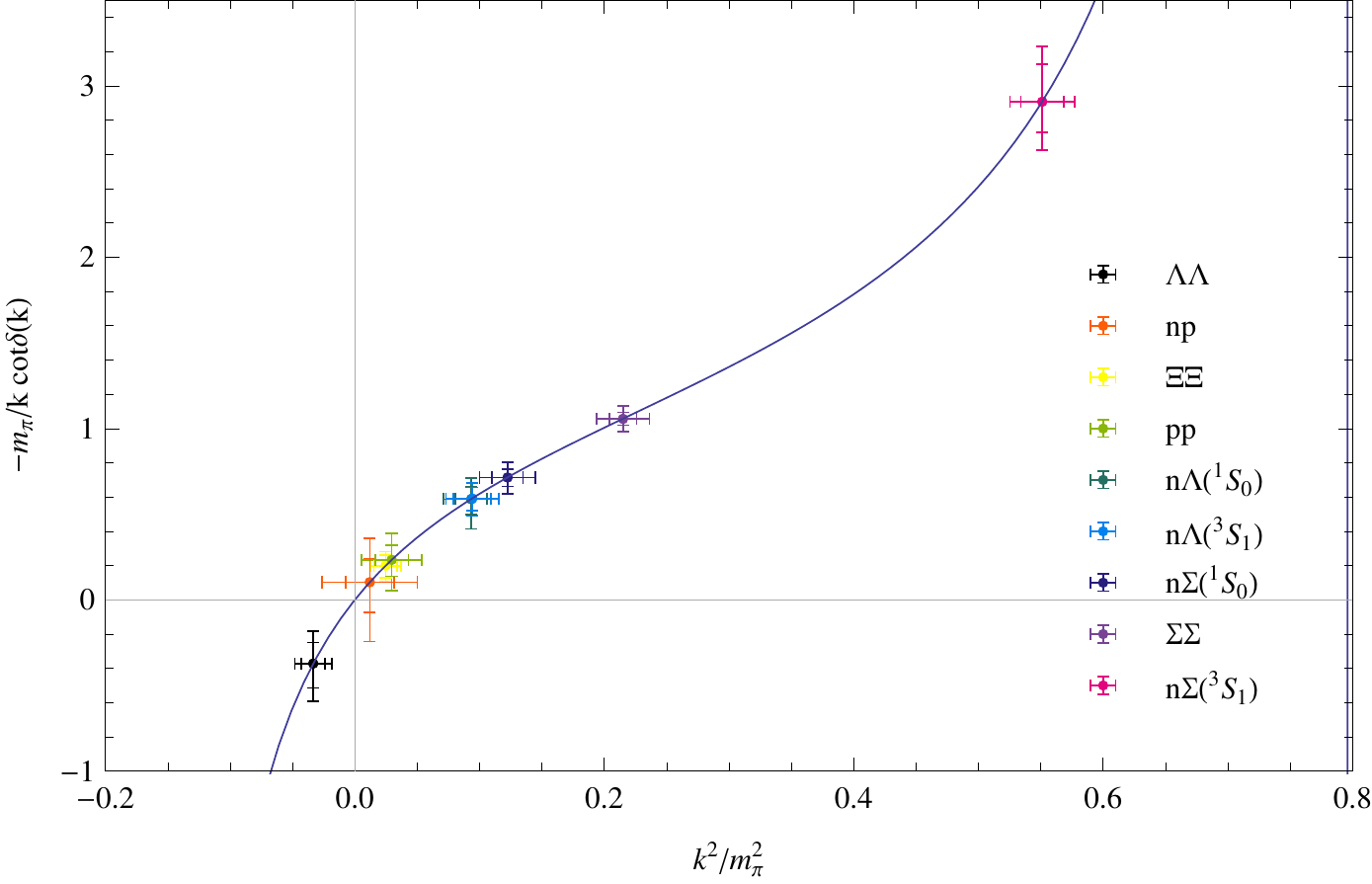}
  \caption{The most recent calculation~\protect~\cite{Beane:2009py} 
of baryon-baryon scattering phase-shifts
    at a pion mass of $m_\pi\sim 390~{\rm MeV}$, calculated on lattices 
generated by
the Hadron Spectrum Collaboration~\cite{Edwards:2008ja,Lin:2008pr}
with a 
spatial-lattice-spacing of $a\sim 0.123~{\rm fm}$, 
lattice-volume of $20^3\times 128$ with an
anisotropy factor of $\xi_T=3.5$.
}
\label{fig:BB}
\end{figure}

\begin{figure}[!ht]
  \includegraphics[height=0.24\textheight]{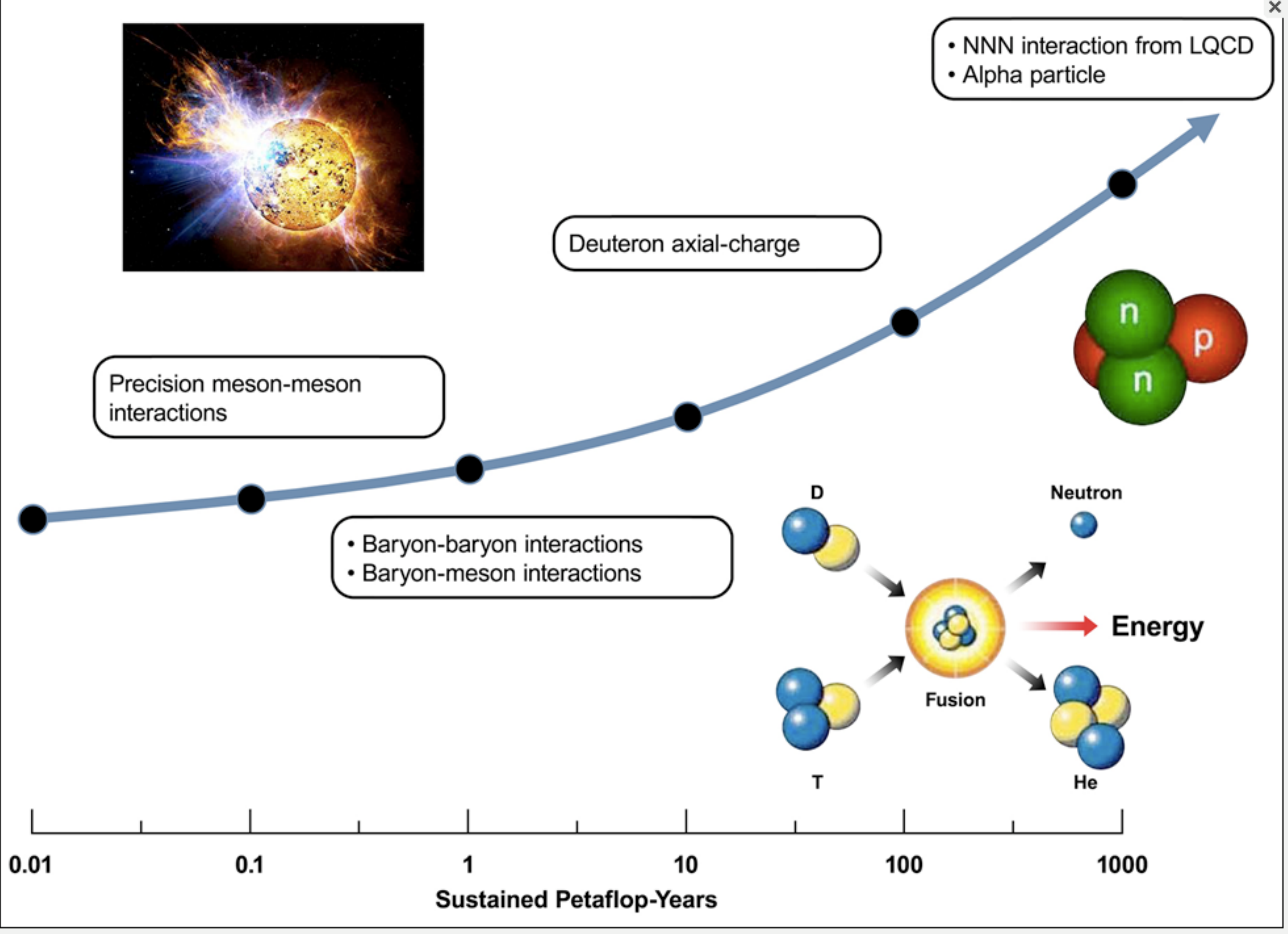}
  \caption{
Estimates of the computational resources required to calculate
the interactions between hadrons~\protect\cite{ExascaleReport:2009}.
}
\label{fig:BBres}
\end{figure}

\begin{figure}[!ht]
  \includegraphics[height=0.11\textheight]{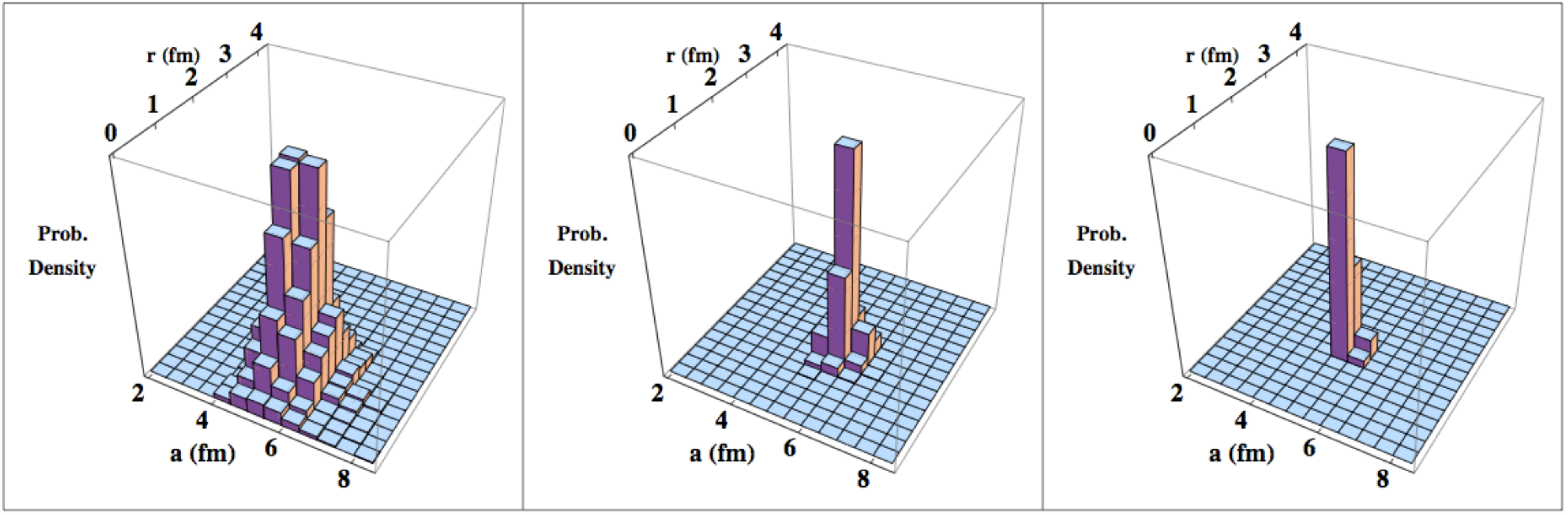}
  \caption{Nucleon-nucleon scattering parameters reconstructed from simulated LQCD
    calculations~\cite{Beane:2010em}. The left panel corresponds to $10\%$
    precision, 
the middle panel to $5\%$
precision and the right panel to $1\%$ precision calculations of the
energy-splitting between two interacting nucleons and two free nucleons.
}
\label{fig:arSim}
\end{figure}

\begin{figure}[!ht]
  \includegraphics[height=0.07\textheight]{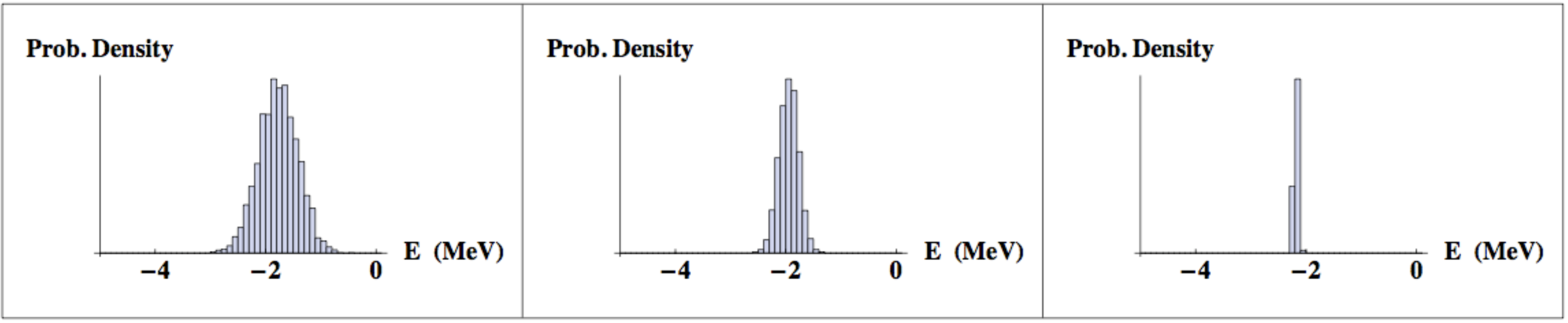}
  \caption{The deuteron binding energy reconstructed from simulated LQCD
    calculations~\cite{Beane:2010em}. The left panel corresponds to $10\%$
    precision, 
the middle panel to $5\%$
precision and the right panel to $1\%$ precision calculations of the
energy-splitting between two interacting nucleons and two free nucleons.
}
\label{fig:DeutSim}
\end{figure}

\begin{figure}[!ht]
  \includegraphics[height=0.28\textheight]{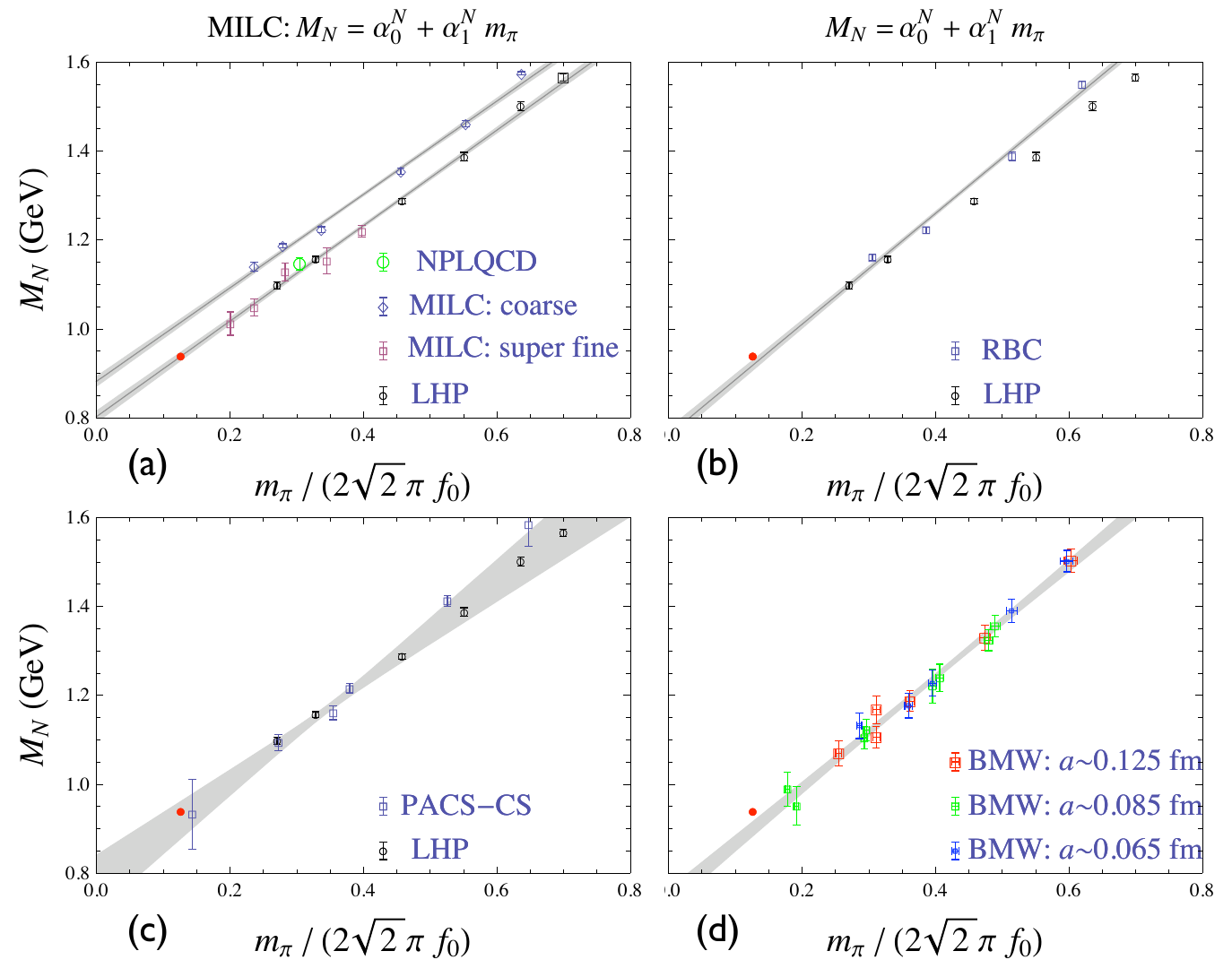}
  \caption{A summary of LQCD calculations of the mass of the 
nucleon)~\protect\cite{WalkerLoud:2008pj}.
I thank A. Walker-Loud for allowing me to reproduce this figure.
}
\label{fig:MN}
\end{figure}

When it comes to calculating the properties of the light nuclei and their
interactions with LQCD, 
the  first ``benchmarking'' step is for  LQCD to
post-dict the large nucleon-nucleon scattering lengths and 
the existence of the deuteron, the simplest nucleus.  
This will firmly establish the 
connection between QCD and nuclear physics, 
and will allow for an exploration of how nuclei and
nuclear interactions depend upon the fundamental parameters of
nature. In particular, it is believed that an understanding of the
fine-tunings that permeate nuclear physics will finally be translated
into fine-tunings of the light-quark masses. While these
issues are of great interest, and it is important to recover what is
known experimentally to high precision, these goals are not the main
objective of the LQCD effort in nuclear physics.  The primary
reason for investing resources in this area
is to be able to calculate physical quantities of importance that
cannot be accessed experimentally, or which can be measured with only
limited precision in the laboratory. Two important examples of how
LQCD calculations can impact nuclear physics are in the
structure of nuclei and in the behavior of hadronic matter at 
densities beyond that of nuclear matter.
LQCD will be able to calculate the
interactions of multiple nucleons, bound or unbound, in the same way that it
can be used to determine the two-body scattering parameters.  
While it is not possible to extract objects from LQCD
calculations  that can be directly
compared with the modern nucleon-nucleon potentials 
(despite claims to the contrary), the scattering-phase shifts
can be determined in the same way that experimentally measured cross-sections
are translated into  phase-shifts.  LQCD calculations of three-nucleon systems
will allow for the determination of the three-nucleon interactions.
Steps towards this goal have been taken with analogous calculations in
meson-systems, e.g. Ref.~\cite{Beane:2007es}, and in the first calculations of
three- and four-nucleon systems~\cite{Beane:2009gs,Yamazaki:2009ua}.  
Two decades of effort in the
development of low-energy effective field theories (EFT's) for nuclear physics 
provide a framework with which to classify the contributions from multi-nucleon
operators, and LQCD calculations will first match onto such operators.
This will provide some sort of unification for nuclear physics as the tools will
then be in place, including those used to perform nuclear structure
calculations starting from modern nucleon-nucleon potentials, to start from QCD
and calculate the properties of multi-nucleon systems, a long-standing goal of
the field.  The results of a recent calculation of baryon-baryon 
phase-shifts~\cite{Beane:2009py}
by the NPLQCD collaboration can be seen in figure~\ref{fig:BB}, and the estimated 
resource requirements to provide precise calculations at the physical pion mass
are shown in figure~\ref{fig:BBres}.
Simulations of LQCD calculations have been performed~\cite{Beane:2010em} to determine the level of
precision required to extract the deuteron binding energy, 
the results of which can be seen in
figures~\ref{fig:arSim} and \ref{fig:DeutSim}.
\vskip 0.1in
\vskip 0.1in
\vskip 0.1in
\vskip 0.1in
In addition to the large computational resources and algorithmic
improvements that are required to complete the mission, 
formal developments are also required. 
For instance, a reliable method with which to extract inelastic
scattering cross-sections from LQCD calculations does not yet exist, and 
must be developed.
Further, a better understanding of the convergence of EFT's that are 
used in such systems is required.  Current LQCD calculations indicate 
unexpected
behaviors in the convergence patterns of the EFT's, including chiral 
perturbation
theory ($\chi$PT). 
Two striking examples of such behaviors are evident in LQCD calculations of the
nucleon mass, exhibiting a light-quark mass dependence that is consistent with 
being linear in the pion mass (i.e. $\sqrt{m_q}$)~\cite{WalkerLoud:2008pj}, 
as shown in figure~\ref{fig:MN},
and in the meson-meson scattering lengths which are consistent with their 
tree-level
predictions, even at large light-quark masses~\cite{Beane:2007xs}.
\vskip 0.1in

\begin{figure}[!ht]
  \includegraphics[height=0.275\textheight]{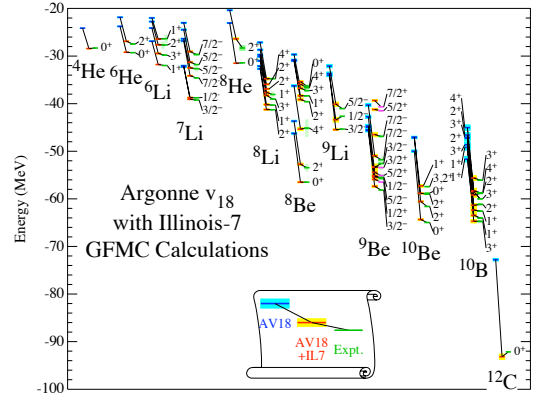}
  \caption{The spectrum of $A=6-12$ nuclei calculated with
    GFMC~\cite{Pieper:2004qw}.
I thank the authors for allowing me to reproduce this figure.
}
\label{fig:GFMC}
\end{figure}

\begin{figure}[!ht]
  \includegraphics[height=0.23\textheight]{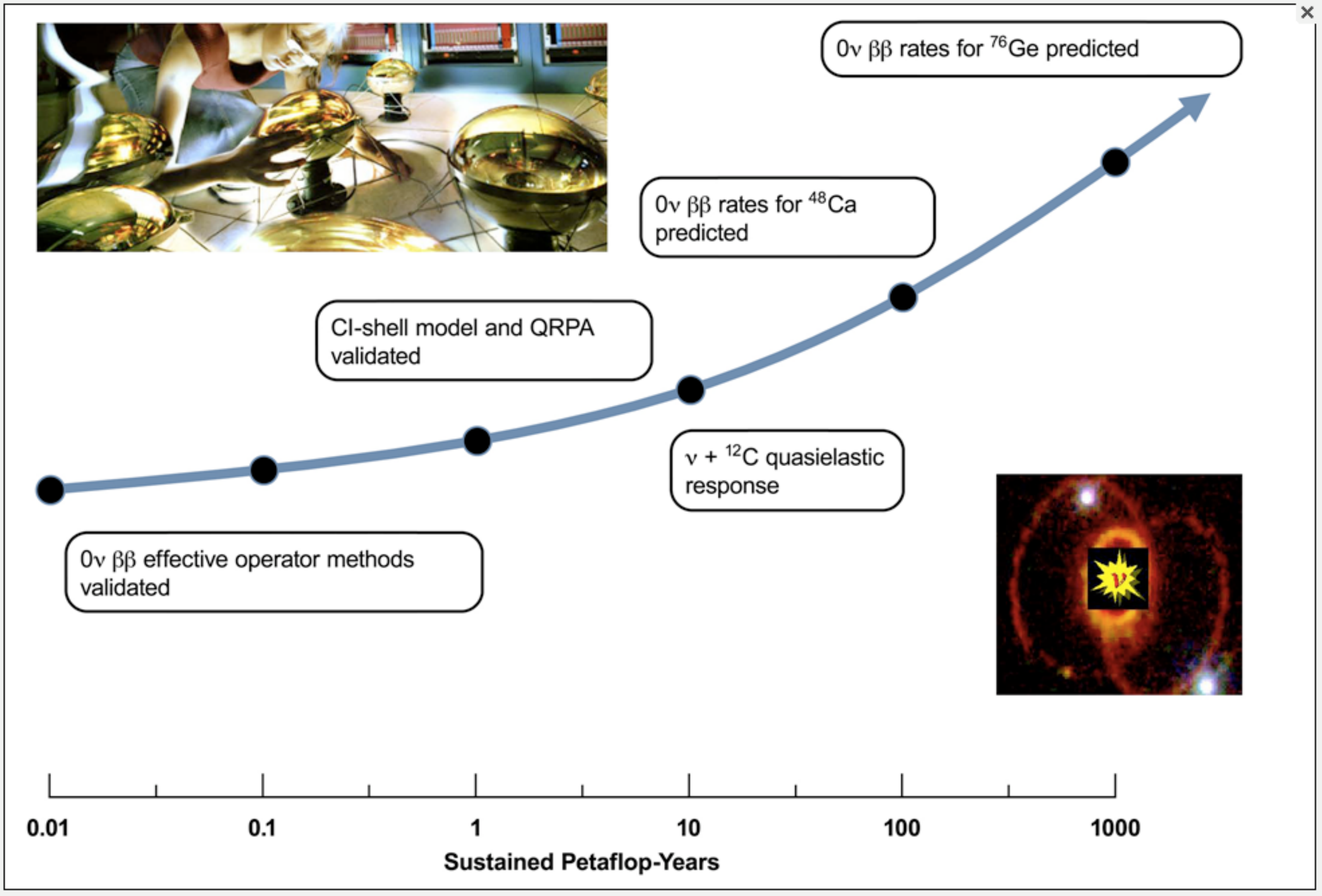}
  \caption{Estimates of the computational resources required to calculate the structure of
    nuclei required to extract neutrino properties from experimental measurements.
}
\label{fig:neutrino}
\end{figure}

While the nuclear structure community has developed impressive tools to perform
calculations starting from either effective interactions in truncated
model-spaces, or in fully no-core-shell-model (NCSM), large computational
resources are required for such calculations.  
Like QCD, this involves  quantum many-body systems, and
exa-scale resources are required to perform calculations of nuclei starting
from the two-nucleon and multi-nucleon forces
beyond the lightest few nuclei.  An example of the
state-of-the-art in the calculation of light nuclei starting from the
nucleon-nucleon and the three-nucleon interaction~\cite{Pieper:2004qw} 
can be seen in figure~\ref{fig:GFMC}
in which the role played by three-body nuclear forces is made clear.
In addition to being extremely important for predictive capabilities for
nuclear processes in environments where experiments are impossible, such
calculations will define how nuclei and their interactions depend upon the
fundamental parameters of nature, namely the strong interaction length-scale,
$\Lambda_{\rm QCD}$, the quark masses, $m_q$, and the electroweak interactions.
The fine-tuning(s) required to produce sufficient carbon for us to
exist
will be translated into fine-tunings of the fundamental
parameters.  Of course, this requires calculating the structure of
alpha-cluster states in
$^8$Be and $^{12}$C, which are
notoriously difficult starting from single particle-orbit type
calculations.  
Such calculations are a
major goal of the nuclear structure community for the next decade, and 
preliminary calculations are underway.
Nuclear structure calculations play an important role in determining the
properties and interactions of neutrinos from present and future neutrino experiments.
An estimate of the computational resources required to perform the nuclear
structure calculations required to  analyze the results of neutrino
experiments are shown in figure~\ref{fig:neutrino}.
\vskip 0.1in
\vskip 0.1in

\begin{figure}[!ht]
  \includegraphics[height=0.23\textheight]{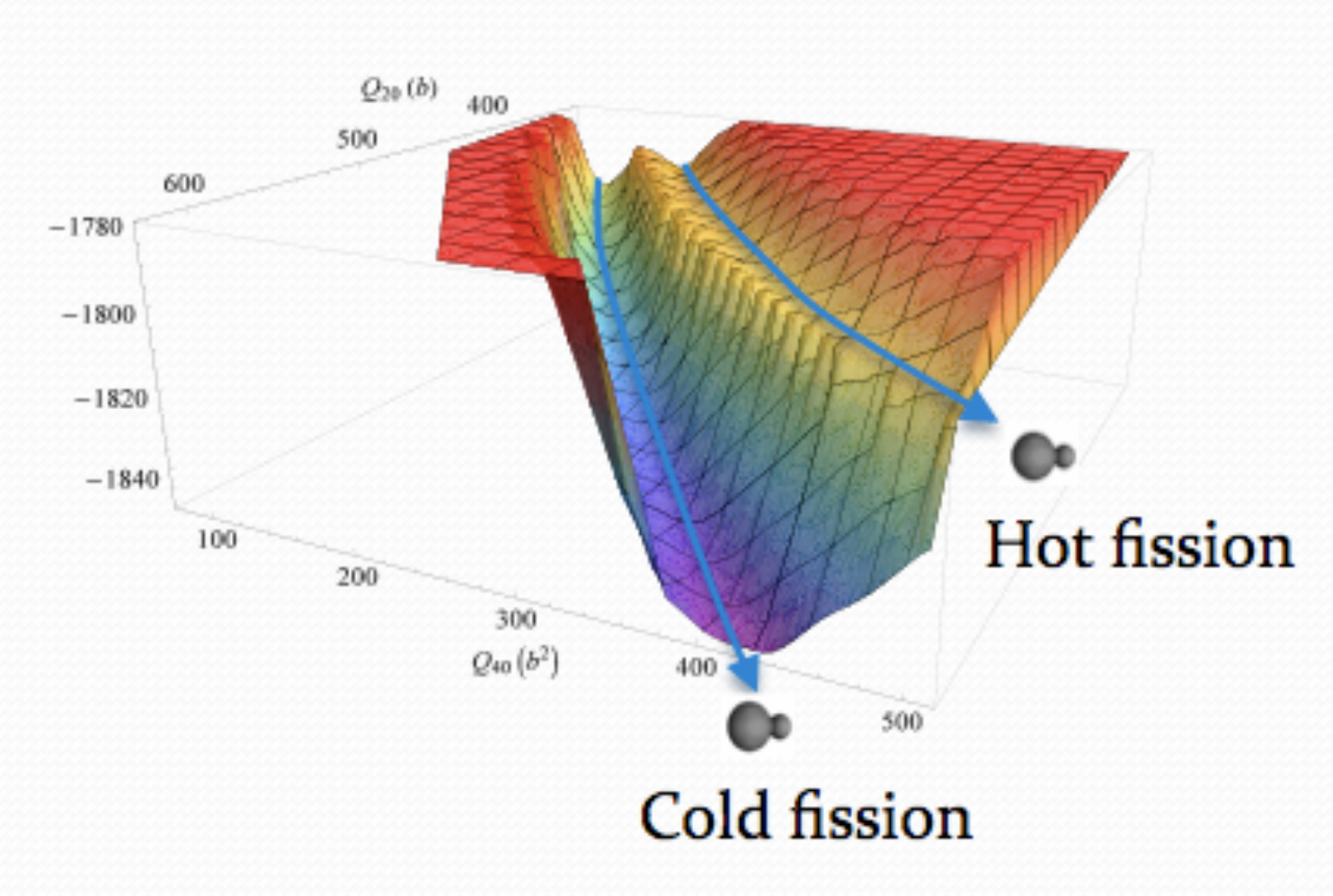}
  \caption{An example of an energy-surface that is found in the fission of a
    heavy-nucleus as function of shape-deformations $Q_{20}$ and $Q_{40}$.
I thank  W. Younes for allowing me to reproduce this figure.
}
\label{fig:fission}
\end{figure}

The nature of calculations in heavy nuclei is quite different to those in
lighter nuclei.
It is not presently feasible to start from the measured or
calculated nucleon-nucleon interaction and the three-nucleon interaction and
calculate, for instance, fission cross-sections directly. 
Calculations starting from an energy density functional have been a focus 
in the community for a number of years, e.g. Ref.~\cite{Bulgac:2010xf}. 
Hartree-Fock-Bogoliubov calculations~\cite{Younes:2009up} 
will require exa-scale
computing to map out the energy-surface associated with the fission
of heavy nuclei.  An example of which 
is shown in figure~\ref{fig:fission},
where it is clear that both cold and hot fission can occur.
A multi-dimensional grid in the parameters describing the nuclear shape,
$Q_{20}$, $Q_{22}$, $Q_{30}$, $Q_{40}$, 
and scission of the nucleus $Q_{N}$, 
requires $\sim 3\times 10^8$ points, and requires an exa-flop
to execute one calculation in less than one day 
(see Ref.~\cite{TL:sciaps10} for a more complete discussion).
\vskip 0.1in

\begin{figure}[!ht]
  \includegraphics[height=0.1\textheight]{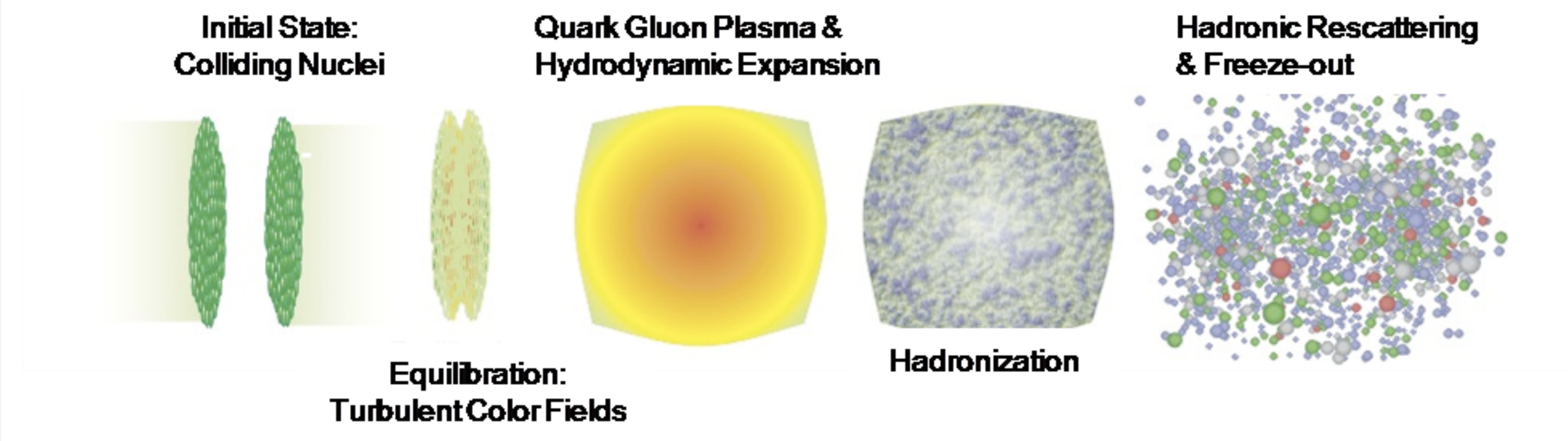}
  \caption{A cartoon of the evolution of a heavy-ion collision.
}
\label{fig:heavyions}
\end{figure}

\begin{figure}[!ht]
  \includegraphics[height=0.29\textheight]{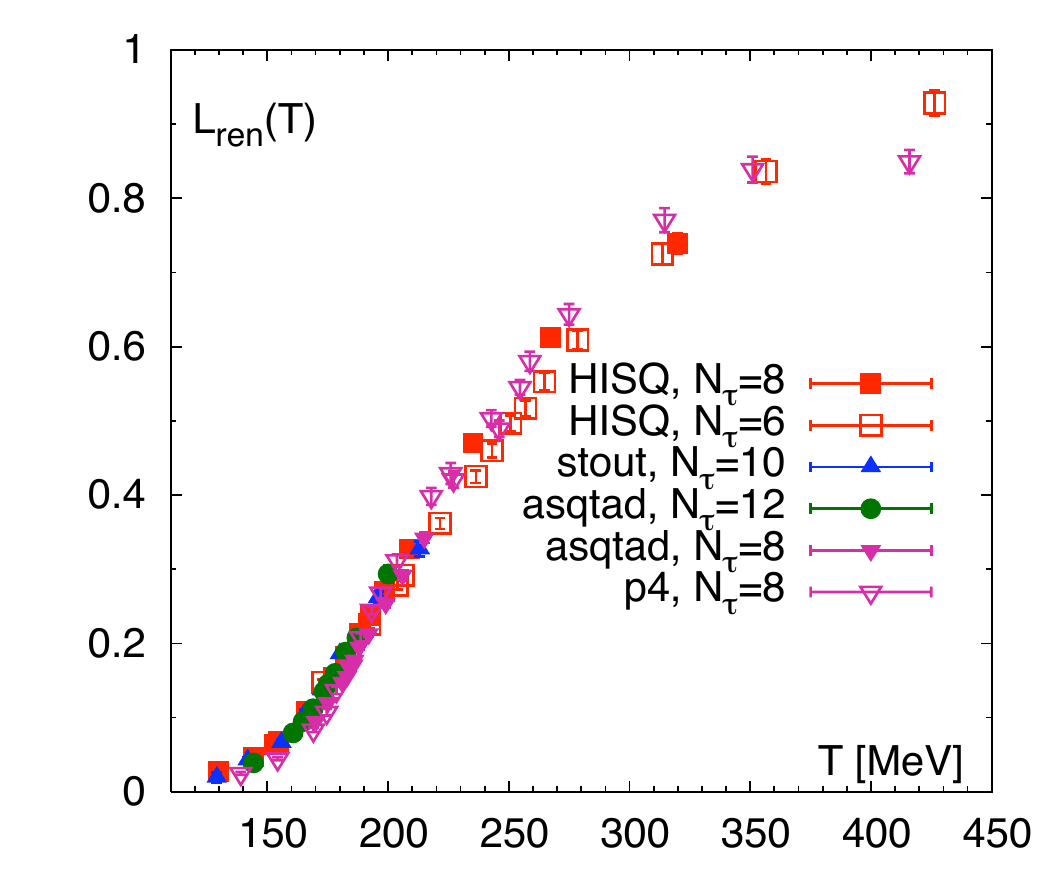}
  \caption{The Polyakov loop as a function of temperature calculated with LQCD
    with a variety of quark discretizations~\protect\cite{Bazavov:2010sb}.
I thank the authors for allowing me to reproduce this figure.
}
\label{fig:Polyakov}
\end{figure}

\begin{figure}[!ht]
  \includegraphics[height=0.25\textheight]{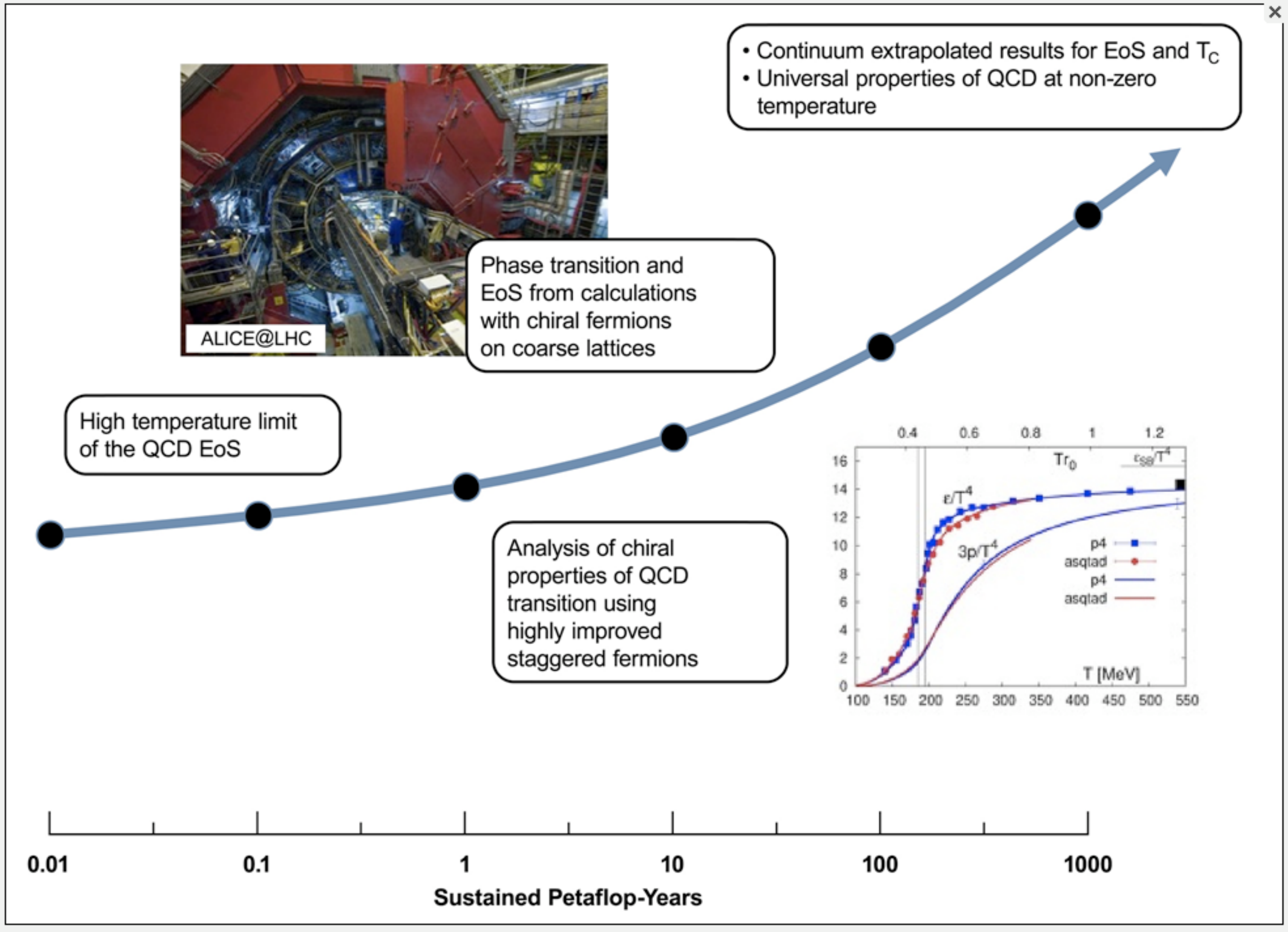}
  \caption{Estimates of the computational resources required to calculate 
the QCD phase-diagram
    at zero baryon number density.
}
\label{fig:hotqcd}
\end{figure}

Exa-scale computing resources are required in order to determine the behavior
of matter under extreme conditions of temperature and density.  Enormous effort
has already been placed into the exploration of equilibrium properties of hot
matter, and to determining the behavior of matter during the collisions of
relativistic heavy-ions (a cartoon of which is shown in
figure~\ref{fig:heavyions}).  
LQCD is the tool with which to calculate
high-temperature properties of systems in equilibrium (see, for example, 
Ref.~\cite{Cheng:2009zi}), 
while the evolution of heavy-ion collisions
requires a host of techniques, some of which are not yet in place.
LQCD has established that the transition from confining to deconfining
at vanishing baryon number density is a cross-over 
at the physical values of the light-quark masses, 
as opposed to, say, a first-order transition, and significant effort is
being placed on mapping out the transition in a number of observables with
different quark discretizations, an example of
which~\cite{Bazavov:2010sb,Bazavov:2009zn} 
is the Polyakov-loop, as shown in figure~\ref{fig:Polyakov}.
A significant conceptual problem that remains in this area is the inclusion of
moderate and large
baryon number density due to the infamous {\it sign-problem} that is
encountered in numerical sampling of the partition function.  
Estimates of the computational
resources required to map out the QCD phase-diagram at zero-baryon-number
density are shown in figure~\ref{fig:hotqcd}.
\vskip 0.1in
\vskip 0.1in
\vskip 0.1in

The other areas of nuclear physics that have been identified as
requiring exa-scale computing
resources to accomplish their goals are astrophysics and accelerator 
design~\cite{ExascaleReport:2009}.
These areas are beyond the scope of this meeting and I will  not discuss 
them further.

Nuclear physicists will need enhanced
collaboration with applied Mathematicians, Computer Scientists, Statisticians,
and others in order to prepare for, and to fully utilize, exa-scale computing
resources when they become available - which is anticipated to 
occur around
2017.
Feedback has already been provided to those designing the
machines by the community, and this will have to be an ongoing dialogue.  
As the nature of the
machines will be quite different from those of today, the codes and
algorithms will need to evolve accordingly.  The codes will have to become
fault-tolerant - they will need to be able to 
survive the loss of multiple  compute-cores.
Further, as the
memory-per-core will likely be substantially less than in today's machines, the
mode of calculation will have to change.  The movement of information is a
major  consumer of power and will need to be minimized.
Consequently, much of the
post-processing, that is done today on smaller machines,
will have to be done {\it in situ}.
In many fields, visualization is crucial to exploring the massive outputs from
calculations.  Nuclear and particle physics is somewhat lagging in this area -
a situation that will have to change within the next few years.
Given the scale of the computational facilities, one should view these
numerical efforts like experiments in their style of operation.
As the nuclear
physics community could not efficiently use an exa-scale 
machine today, but could
efficiently use a 1-10 sustained peta-flop machine,
a staged
evolution to the exa-scale will likely be  required.
However, it must be stressed that computing power is insufficient by itself to
accomplish all of the goals.  
Conceptual and formal developments must take place in some areas, hand-in-hand,
with the deployment of computing resources 
and developments in 
algorithms and applied mathematics.

The field of nuclear physics will be somewhat revolutionized by the deployment
of exa-scale computing resources.
Exa-scale computing will provide nuclear physics with predictive capabilities
that allow for the assignment of reliable uncertainty estimates in
observables that cannot be explored experimentally.
Further, it will enable the systematic exploration of
fundamental aspects of nature that are manifested in
the structure and interactions of nuclei.

\begin{theacknowledgments}
I would like to thank all those who
participated in the {\it Scientific Grand Challenges Workshop Series}, and my
collaborators in NPLQCD.  NT@UW-10-25.
\end{theacknowledgments}

\end{document}